\documentclass[paper]{agujournal2019}
\usepackage{url}
\usepackage{lineno}
\usepackage{color}
\usepackage{amsmath,amsfonts,amssymb}
\usepackage{comment}
\usepackage{subcaption}
\draftfalse

\newcommand{\ssr}{   {Space Sci. Rev. }}

\newcommand{\jgr}{   {J. Geophys. Res.}}
\newcommand{\grl}{   {Geophys. Res. Lett.}}

\newcommand{\prl}{   {Phus. Rev. Lett.}}

\graphicspath{{./}{figures/}}

\journalname{JGR: Space Physics}

\begin{document}

%% ---------------------------------------------------------------%%

\title{Properties of Intense Electromagnetic Ion Cyclotron Waves: Implications for Quasi-linear, Nonlinear, and Nonresonant Wave-Particle Interactions}

\authors{Xiaofei Shi\affil{1}, Anton Artemyev\affil{1}, Xiao-Jia Zhang\affil{2,1}, Didier Mourenas\affil{3,4}, Xin An\affil{1},  Vassilis Angelopoulos\affil{1}} 
\affiliation{1}{Department of Earth, Planetary, and Space Sciences, University of California, Los Angeles, Los Angeles, CA, 90095, USA}
%\affiliation{2}{Space Research Institute of the Russian Academy of Sciences, Moscow, 117997, Russia}
\affiliation{2}{Department of Physics, University of Texas at Dallas, Richardson, TX, USA}
\affiliation{3}{CEA, DAM, DIF, Arpajon, France}
\affiliation{4}{Laboratoire Matière en Conditions Extrêmes, Université Paris-Saclay, CEA, Bruyères-le-Châtel, France}

\correspondingauthor{Xiaofei Shi}{sxf1698@g.ucla.edu}

\begin{abstract}
Resonant interactions between relativistic electrons and electromagnetic ion cyclotron (EMIC) waves provide an effective loss mechanism for this important electron population in the outer radiation belt. The diffusive regime of electron scattering and loss has been well incorporated into radiation belt models within the framework of the quasi-linear diffusion theory, whereas the nonlinear regime has been mostly studied with test particle simulations. There is also a less investigated, nonresonant regime of electron scattering by EMIC waves. All three regimes should be present, depending on the EMIC waves and ambient plasma properties, but the occurrence rates of these regimes have not been previously quantified. This study provides a statistical investigation of the most important EMIC wave-packet characteristics for the diffusive, nonlinear, and nonresonant regimes of electron scattering. We utilize 3 years of Van Allen Probe observations to derive distributions of wave amplitudes, wave-packet sizes, and rates of frequency variations within individual wave-packets. We demonstrate that EMIC waves typically propagate as wave-packets with $\sim 10$ wave periods each, and that $\sim 3-10$\% of such wave-packets can reach the regime of nonlinear resonant interaction with 2 to 6 MeV electrons. We show that EMIC frequency variations within wave-packets reach $50-100$\% of the center frequency, corresponding to a significant high-frequency tail in their wave power spectrum. We explore the consequences of these wave-packet characteristics for high and low energy electron precipitation by H-band EMIC waves and for the relative importance of quasi-linear and nonlinear regimes of wave-particle interactions.
\end{abstract}

\begin{keypoints}
\item Most intense H-band EMIC wave packets are short but 3-10\% may lead to nonlinear interactions
\item The power spectrum of intense H-band EMIC wave packets contains a significant high-frequency tail
\item Resonant interactions with intense H-band EMIC wave packets likely play an important role in sub-relativistic electron precipitation
\end{keypoints}

\section{Introduction}\label{sec:intro}
Electromagnetic ion cyclotron (EMIC) waves are one of the main waves controlling the dynamics of relativistic electron fluxes via resonant wave-particle interactions \cite<see reviewers by>[ and references therein]{Millan&Thorne07,Shprits08:JASTP_local,Usanova21:frontiers}. These waves are generated by hot transversely anisotropic ions \cite{Sagdeev&Shafranov61,Thorne&Kennel71} that are either injected from the night-side plasma sheet or heated by day-side magnetosphere compressions \cite<see EMIC waves statistics in>{Usanova12,Zhang16:grl,Jun19:emic,Jun21:emic}.

\subsection{Quasi-linear electron resonant interaction with EMIC waves}

The classical and most widespread treatment of electron scattering by EMIC waves is quasi-linear diffusion through cyclotron resonance, which depends on the time-averaged wave power as a function of the wave frequency normalized to the proton gyrofrequency \cite<e.g.,>{Thorne&Kennel71, Albert03, Summers&Thorne03, Kersten14,Ni15,Shprits16, Drozdov17,Ross20}. However, this approach may not be sufficient to explain observations of sub-relativistic electron precipitation below $\sim 0.5$ MeV during conjunctions with EMIC wave bursts near the equator \cite{Hendry17, Capannolo19, An22:prl,Angelopoulos23:SSR}, because the frequency of the most intense observed EMIC waves is usually not sufficiently high to reach cyclotron resonance with sub-relativistic electrons \cite{Kersten14,Ni15}. Moreover, the quasi-linear approach cannot describe a potentially important regime of wave-particle interactions: nonlinear resonant interactions \cite<see examples in>{Albert&Bortnik09,Omura&Zhao12,Omura&Zhao13,Grach&Demekhov18:I,Grach&Demekhov18:II}.
 
\subsection{Nonlinear electron resonant interaction with EMIC waves}

EMIC wave amplitudes often exceed hundreds of pT \cite{Meredith14, Zhang16:grl} and are large enough to reach the regime of nonlinear resonant interaction, when the wave Lorentz force exceeds the electron mirror force in an inhomogeneous magnetic field, and thus significantly changes electron trajectories \cite<see details of the basic concept of nonlinear resonant interaction in reviews by>{Omura91:review,Shklyar09:review,Albert13:AGU,Artemyev18:cnsns}. Such a significant effect on electron dynamics breaks the key assumption of the quasi-linear theory which is the approximation of unperturbed particle trajectories and requires consideration of more sophisticated effects than diffusive scattering \cite{Albert&Bortnik09}. Quite comprehensive parametric investigations of possible nonlinear resonant effects for electron interactions with EMIC waves can be found in \citeA{Kubota15,Kubota&Omura17,Grach&Demekhov20,Hanzelka23:emic}. Here, we discuss the two most important nonlinear processes: phase bunching and phase trapping. 

Electron phase bunching is a nonlinear resonant effect occurring when the wave field is sufficiently strong to keep electrons in resonance over a time scale of multiple gyroperiods, changing their pitch-angle (and energy). Such phase bunching by EMIC waves results in an increase in pitch-angle, and is not a diffusive process: an ensemble of electrons with the same energy and pitch-angle, but random gyrophases, will experience pitch-angle increase with a finite mean value \cite<see examples in>{Albert&Bortnik09,Su12:emic,Grach&Demekhov18:II,Grach&Demekhov20}. Such pitch-angle drift in phase space moves electrons away from the loss-cone and can potentially reduce the efficiency of electron precipitation \cite<see discussion in>{Bortnik22:grl}. The pitch-angle change due to a single resonant interaction (during one bounce period) scales with the normalized amplitude $B_w/B_0$ of EMIC waves as $\Delta\alpha\propto\sqrt{B_w/B_0}$ \cite<e.g.,>[and references therein]{Albert&Bortnik09}, and is larger than the drift due to diffusion $\Delta\alpha\propto (B_w/B_0)^2$ \cite<see discussion in>{Bortnik22:grl}. Phase bunching is a local process and only depends on wave amplitude at the resonance. Therefore, the important EMIC wave characteristic that is required to quantify this process is the range of variation of instantaneous wave intensity. Note that the existing EMIC wave datasets \cite{Meredith14, Zhang16:grl, Ross21} provide only time-averaged wave power from Fourier analysis of EMIC spectra which, for short wave-packets, can lead to a significant underestimation of the instantaneous peak wave intensity. 

Electron phase trapping is a nonlinear resonant effect occurring when the wave field can lock electrons around the resonance on a time-scale of a fraction of an electron bounce period, significantly changing their pitch-angles (and energies). Such phase trapping by EMIC waves results in a decrease in pitch-angle, and constitutes one of the main mechanisms responsible for fast electron precipitation by intense EMIC waves \cite{Kubota&Omura17,Grach21:emic,Hanzelka23:emic}. Such trapping is nonlocal, i.e., the efficiency of the associated electron transport in pitch-angle and energy depends on the shape/size of EMIC wave-packets and their coherence \cite<see discussion of this effect for various waves types in>{Tao13,Mourenas18:jgr,Zhang20:grl:phase,Grach&Demekhov20,An22:Tao}. Therefore, for an accurate quantification of phase trapping we need information on the fine structure of EMIC wave-packets, including wave-packet size, wave phase, and frequency variations that may destroy phase trapping. 

\subsection{Nonresonant electron resonant interaction with EMIC waves}\label{Sec:NR}

In contrast to the nonlinear resonant interaction, which is responsible for rapid electron flux variations, nonresonant effects \cite{Chen16:nonresonant} are relatively weak, second-order effects. However, these nonresonant effects can play a crucial role in the precipitation of sub-relativistic electrons. Such sub-relativistic precipitation associated with EMIC waves has been frequently observed \cite<e.g.,>{Capannolo19:microburst,Angelopoulos23:SSR}. However, it cannot be described by resonant interactions with the most intense observed EMIC waves, because such resonant interactions are often limited to energies above $1$ MeV, due to the insufficiently high frequency of the main waves \cite<e.g.,>{Kersten14,Ni15,Ross21}. Although some part of the sub-relativistic precipitation could perhaps be explained by hot plasma effects on the EMIC wave dispersion \cite{Silin11,Bashir22:grl}, such hot plasma effects usually tend to increase the minimum resonant energy and cannot help sub-relativistic precipitation \cite{Cao17,Chen19}. One well-known type of nonresonant electron scattering can be understood as some sort of curvature scattering \cite{bookChirikov87,Birmingham84,Buchner89, Delcourt94:scattering,Artemyev15:jgr:butterfly}, where the pitch-angle change is exponentially scaled with its distance in energy from the resonance energy \cite<see general theory in>{Neishtadt00}. In the case of EMIC waves, the wave-packet size, or the wave amplitude modulation, plays a crucial role in the determination of the wave number range (i.e., the range of spatial scales of magnetic field fluctuations) for the nonresonant interaction and, thus, controls the efficiency of this effect \cite{Chen16:nonresonant,An22:prl,An23:arXiv}.

Over a short spatial extent, an EMIC wave-packet can contain a broad spectrum of wavenumbers $k$ distributed around the wavenumber of peak power, $k_0$. The upper range of this $k$ spectrum allows cyclotron resonant interactions between electrons of lower-than-typical energy with waves of much smaller amplitude than the main waves, corresponding to the wavenumber $k_0$, which interact with higher energy electrons. Since EMIC wave frequencies $\omega$ are lower than the proton gyrofrequency $\Omega_{ci}$ and much lower than the electron gyrofrequency $\Omega_{ce}=1836\,\Omega_{ci}$, the cyclotron resonance condition for parallel EMIC waves can be rewritten as $kv/\Omega_{ce}=1/(\gamma \cos\alpha)$, with $\gamma$ the Lorentz factor, $v$ and $\alpha$ the electron velocity and pitch-angle \cite{Summers&Thorne03, Angelopoulos23:SSR}. Consequently, any magnetic fluctuation of sufficiently low frequency, $\omega\ll\Omega_{ce}$, can resonantly scatter low energy electrons if its wave number $k$ is sufficiently high to satisfy the above resonance condition \cite<e.g., see>{Xu&Egedal22}. For typical high-$k$ EMIC waves of low amplitudes \cite{Denton19}, this resonant scattering is proportional to the wave power and much more efficient than purely nonresonant scattering \cite{Xu&Egedal22, An22:prl, Angelopoulos23:SSR}. Therefore, the nonresonant electron interactions with EMIC waves could be more precisely recast as nonresonant with the main EMIC waves (at peak wave power) while still resonant with much lower intensity EMIC waves at higher wave numbers $k$, which usually correspond to higher $\omega/\Omega_{ci}$ values based on the EMIC wave dispersion relation \cite{Summers&Thorne03, Denton19}. To quantify the effects of such sub-relativistic electron interactions with EMIC wave-packet edges, we need statistical information about EMIC wave-packet characteristics, such as the wave power spectrum tail at high $\omega/\Omega_{ci}$, which should correspond to high wave numbers $k$.

\subsection{On the most significant characteristics of EMIC wave-packets}

Because of their importance for identifying the different regimes of wave-particle interactions, we statistically investigate the following EMIC wave-packet characteristics: wave-packet size, percentage of wave packets that can interact resonantly with electrons nonlinearly, and details of variations of wave frequency and wave power within wave-packets. We start with a description of the EMIC wave dataset and the methods of wave-packet determination in Sect.~\ref{sec:data}. Then in Sect.~\ref{sec:packets} we describe statistical characteristics of wave-packet amplitudes and sizes, and in Sect.~\ref{sec:frequency} we provide information about the inner structure of wave-packets. Sect.~\ref{sec:Implications} examines the consequences of wave packet characteristics for the energy of precipitating electrons, and the relative importance of quasi-linear, nonlinear, and nonresonant wave-particle interactions. Finally, in Sect.~\ref{sec:conclusions} we summarize the results.

\section{Spacecraft Instruments and Dataset}\label{sec:data}

The twin Van Allen Probes were launched on 2012-08-30 into a near-equatorial elliptical orbit with geocentric apogee $5.8 R_E$ and perigee $1.1 R_E$ \cite[]{Mauk13}. High-resolution magnetic field data are obtained from the Electric and Magnetic Field Instrument Suite and Integrated Science (EMFISIS) fluxgate magnetometer \cite[]{Kletzing13}. The plasma density is derived from the upper hybrid frequency measured by EMFISIS High-Frequency Receiver (HFR) \cite[]{Kurth15} or can be calculated using spacecraft potential from the Electric Field and Waves instrument \cite[]{Wygant13}.  

\begin{figure*}
\centering
\includegraphics[width=1\textwidth]{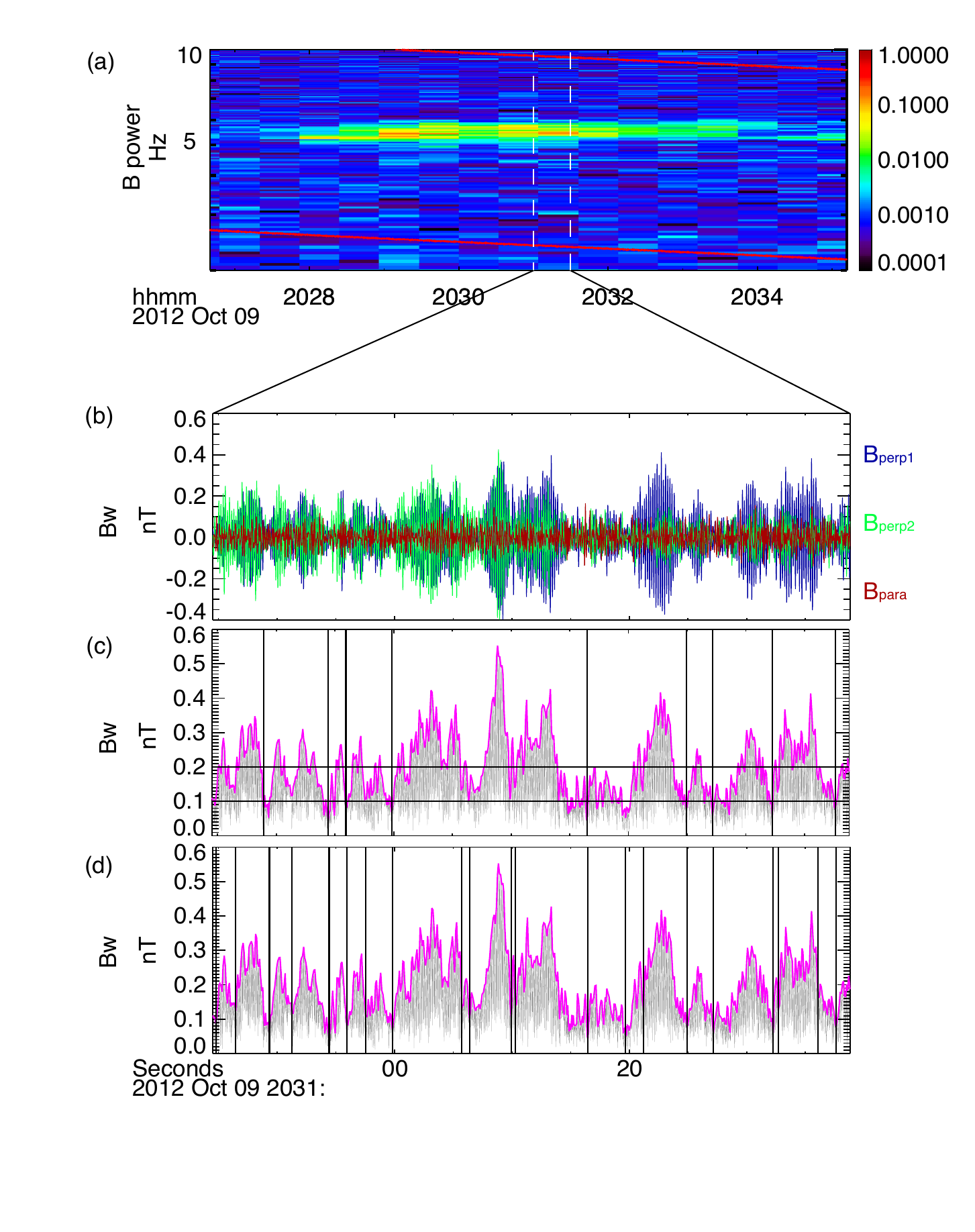}
\caption{Observations of H band EMIC waves: (a) wave power spectrum. Two red lines from top to bottom indicate proton and helium cyclotron frequency, respectively; (b) two perpendicular components and the parallel component of wave magnetic field in field-aligned coordinate; (c) Wave amplitude and wave packets determined by criterion (1); (d) wave packets determined by criterion (2).}
\label{fig1}
\end{figure*}

In this study, we examined proton band EMIC wave events collected by Van Allen Probes from 2012-2015 \cite<the EMIC wave dataset is the same as in>{Zhang16:grl}. To select wave packets in each event, we used two different criteria: (1) wave packets with peak amplitude $B_{w,peak}>200$ pT and dips in amplitude below $100$ pT on each side of the peak; (2) wave packets with peak amplitude $B_w>200$ pT and dips in amplitude below $0.5B_{w,peak}$ on each side of the peak.  Then we calculated the wave packet size ($\beta$) as the number of wave periods between two edges of each packet. The main difference between these two criteria is that when the wave amplitude is large, i.e., $B_{w,peak}>500$ pT, the noticeable dips near the amplitude peak may not be smaller than $100$ pT. Thus the first criterion will favor large packet sizes for intense waves. This bias is avoided by the second criterion in which the threshold for the amplitude dip is not fixed but varies with the peak amplitude, allowing detection of sub-packets when the waves are particularly intense. Figure \ref{fig1} shows an observation of EMIC waves and the wave packets selected using the two above-described criteria. Note that the wave amplitude used to select the packet is the envelope of the full wave amplitude (the magenta lines in Figure \ref{fig1}(c,d)).

\section{Wave-packet statistics}\label{sec:packets}

\begin{figure*}
\centering
\includegraphics[width=1\textwidth]{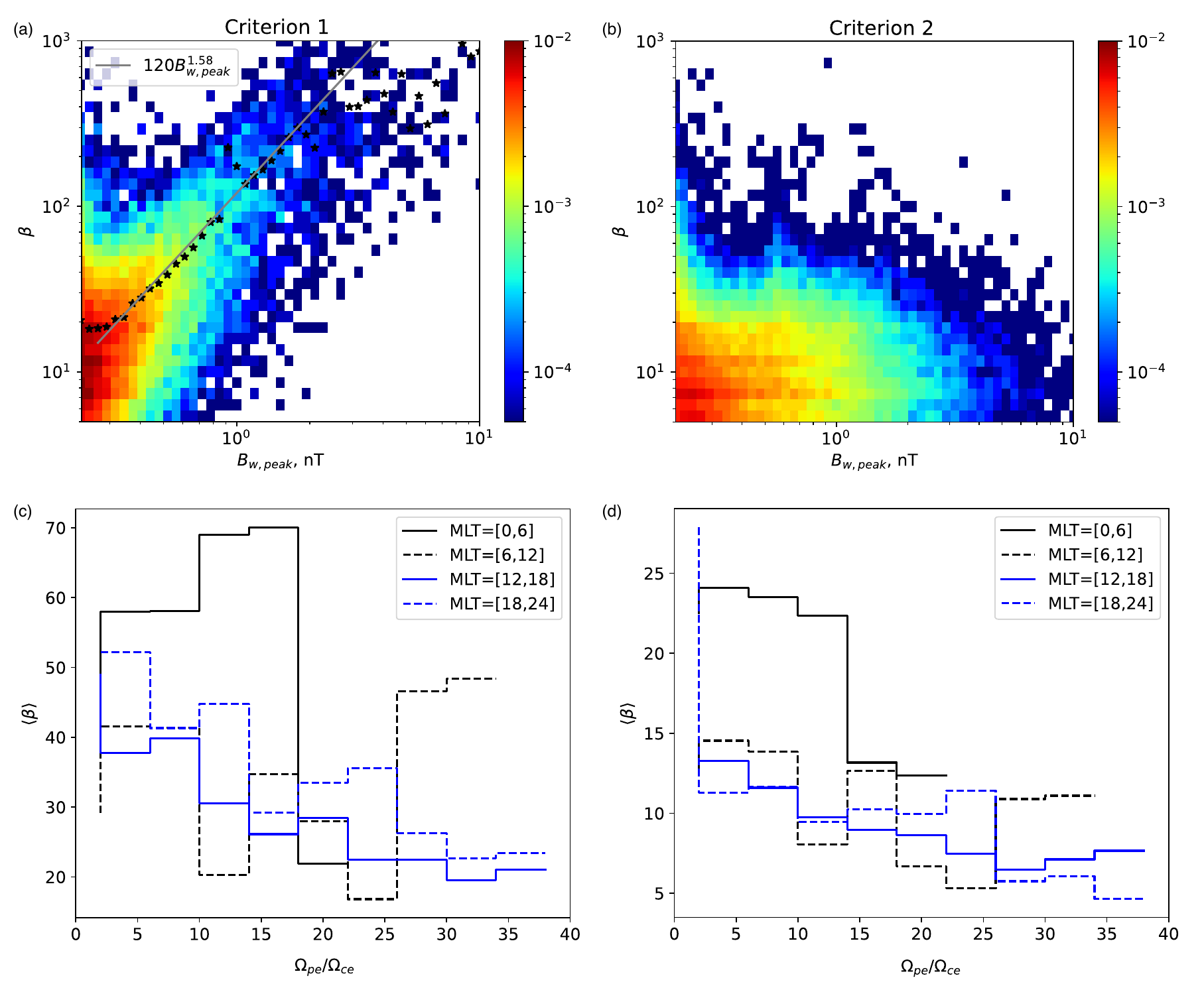}
\caption{Probability distribution of H-band EMIC wave packets in $(\beta, B_w)$ space: (a) results from Criterion 1. Crosses denote the mean $\beta$ for each $B_{w,peak}$ bin and the solid black line shows the least-square fit $\beta = 120\,B_{w,peak}^{1.58}$. (b) Results from Criterion 2. (c) Average $\beta$ as a function of $\Omega_{pe}/\Omega_{ce}$ and $\beta$ in 4 different $MLT$ ranges using Criterion 1. (d) Same as (c) using Criterion 2. }
\label{fig2} 
\end{figure*}

Figures \ref{fig2}(a) and \ref{fig2}(b) show the probability distribution of wave-packet size $\beta$ measured using Criteria 1 and 2 (top-left and top-right, respectively), as a function of peak wave-packet amplitude $B_{w,peak}$. Most wave-packets have packet sizes in the range of $\beta\sim 5-30$. The results using Criterion 1 demonstrate that wave packet size $\beta$ increases with peak amplitude approximately as $\beta\approx 120\,B_{w,peak}^{1.58}$. This trend is roughly consistent with an average wave packet shape $B_w(t)/B_{w,peak}\approx \epsilon/(|t-t_{peak}|^{1/1.58}+\epsilon)$ down to the threshold at $B_w=0.1$ nT, with $\epsilon\ll\max(|t-t_{peak}|^{1/1.58})$, $B_w(t_{peak})=B_{w,peak}$, and wave-packet size $\beta$ determined by the temporal scale $\beta \sim |t-t_{peak}|$. It corresponds to average (and minimum) amplitudes decreasing more and more slowly farther away from $B_{w,peak}$. Figure \ref{fig1}(c) shows several examples of packets with this shape. Such a shape is also in agreement with the much shorter size of these packets at half of their peak amplitude using Criterion 2 in Figure \ref{fig1}(d). A similar distribution of wave packet sizes and amplitudes has been observed for whistler-mode chorus wave packets \cite{Zhang20:grl:frequency, Zhang21:jgr:data&model, Nunn21, Mourenas22:jgr:ELFIN}. Using Criterion 2, the distribution of packet sizes $\beta$ is less dependent on wave amplitude (compare top-left and top-right panels in Fig. \ref{fig2}), which could stem from a different physical origin of the most significant amplitude modulation close to the peak (i.e., likely a nonlinear modulation, perhaps with some wave superposition) compared to other amplitude modulations occurring farther away from the peak.

We also collected the wave frequency, background magnetic field, and ratio of $\Omega_{pe}=2\pi f_{pe}$ (electron plasma frequency) to $\Omega_{ce}=2\pi f_{ce}$ (electron cyclotron frequency) measured simultaneously with each wave-packet. Figure \ref{fig2}(c,d) shows the averaged $\beta$ of H-band EMIC wave packets as a function of $\Omega_{pe}/\Omega_{ce}$ in 4 different $MLT$ ranges. Based on both Criteria 1 and 2, the size $\beta$ of wave-packets at 12-24 MLT decreases by a factor $\approx2$ as $\Omega_{pe}/\Omega_{ce}$ increases from $5$ to $35$. There is no clear dependence of $\beta$ on MLT over 6-24 MLT, whereas $\beta$ is somewhat larger at 0-6 MLT.

To further understand the properties of EMIC waves and the regime of resonant interaction with electrons, we calculated the resonant interaction's magnetic latitude along the field line for EMIC waves and the inhomogeneity parameter $S$ which controls the dynamics of resonant electrons \cite[]{Omura&Zhao12}. These are explained below.

The wave dispersion relation for proton band EMIC waves can be written as:

\begin{equation}
    \left(\frac{kc}{\omega}\right)^2=1-\frac{\Omega_{pe}^2}{\omega(\omega+\Omega_{ce})}-\frac{\Omega_{pi}^2}{\omega(\omega-\Omega_{ci})}
\end{equation}
where $\omega$ is the wave frequency, $k$ is the wave vector, $c$ is the speed of light, $\Omega_{ps}$ and $\Omega_{cs}$ are plasma frequency and cyclotron frequency for species $s$ ($i$ for proton, $e$ for electron), respectively. 

We use a dipole model for the background magnetic field: $B(\lambda)=B_{eq}\sqrt{1+3\sin^2\lambda}/\cos^6\lambda=B_{eq}f(\lambda)$ with $B_{eq}$ the magnetic field strength at the equatorial plane and $\lambda$ the magnetic latitude. Therefore, the electron cyclotron frequency varies with latitude as $\Omega_{ce}=f(\lambda)\Omega_{ce, eq}$. We assume that the plasma frequency $\Omega_{pe}=\sqrt{ne^2/m_e\epsilon_0}$ is a constant along the field line. The resonance condition for relativistic electrons depends on the magnetic latitude:
\begin{equation}
    \omega-k(\lambda)v_{\parallel}(\lambda)=-\frac{\Omega_{ce}}{\gamma}
\end{equation}
where $v_{\parallel}$ is the electron velocity parallel to the background magnetic field and $\gamma$ is the relativistic factor. At the latitude of cyclotron resonance, the parallel velocity of the resonant electron is determined by $\gamma$ and the equatorial pitch angle $\alpha_{eq}$:
\begin{equation}
    v_{\parallel}(\lambda)=c\sqrt{1-\gamma^{-2}}\sqrt{1-f(\lambda)\sin^2\alpha_{eq}}
\end{equation}
Combining equations (1-3) we can get the resonance latitude, $\lambda_R$.
%\begin{equation}
%    \omega-\omega\sqrt{1-\frac{\Omega_{pe}^2}{\omega(\omega+\Omega_{ce,eq}f(\lambda))}-\frac{\Omega_{pi}^2}{\omega(\omega-\Omega_{ci,eq}f(\lambda))}}\sqrt{1-\gamma^{-2}}\sqrt{1-f(\lambda)\sin^2\alpha_{eq}}=-\frac{\Omega_{ce,eq}f(\lambda)}{\gamma}
%\end{equation}
With $\omega$, $\Omega_{ce,eq}$, and $\Omega_{pe, eq}/\Omega_{ce, eq}$ obtained from observation, we can determine the resonance latitude $\lambda (E,\alpha_{eq})$ at different energies ($E=m_ec^2(\gamma-1)$) and $\alpha_{eq}$. Then we calculate the inhomogeneity ratio parameter $S$, defined as \cite{Omura&Zhao12}:
\begin{equation}
    S=-\frac{B/B_w}{kv_{\perp}\Omega_{ce}/\gamma}\left(\frac{\omega}{\Omega_e}\left(\frac{v_{\perp}^2-V_{R}^2}{2V_p^2}+\frac{V_R^2}{V_gV_p}\right)+\frac{V_R}{\gamma V_p}\right)V_p\frac{\partial\Omega_{ce}}{\partial r_\parallel}
\end{equation}
where $V_R=(\omega+\Omega_{ce}/\gamma)/k$, $V_p$ is the wave phase velocity, $V_g$ is wave group velocity, and 
\[\frac{\partial\Omega_{ce}}{\partial r_\parallel}=\frac{\partial f(\lambda)}{\partial \lambda}\frac{\partial \lambda}{\partial r_\parallel}\Omega_{ce,eq}\]
with $r_\parallel$ the direction along the field line. $S$ is a function of resonance latitude. It is proportional to the background field gradient and inversely proportional to the wave intensity. When $|S|<1$, the waves are intense enough to locally overcome the mirror force and alter the electron trajectory significantly, enabling nonlinear interactions. 

\begin{figure*}
\centering
\begin{tabular}{ c c  }
  \includegraphics[width=1\textwidth]{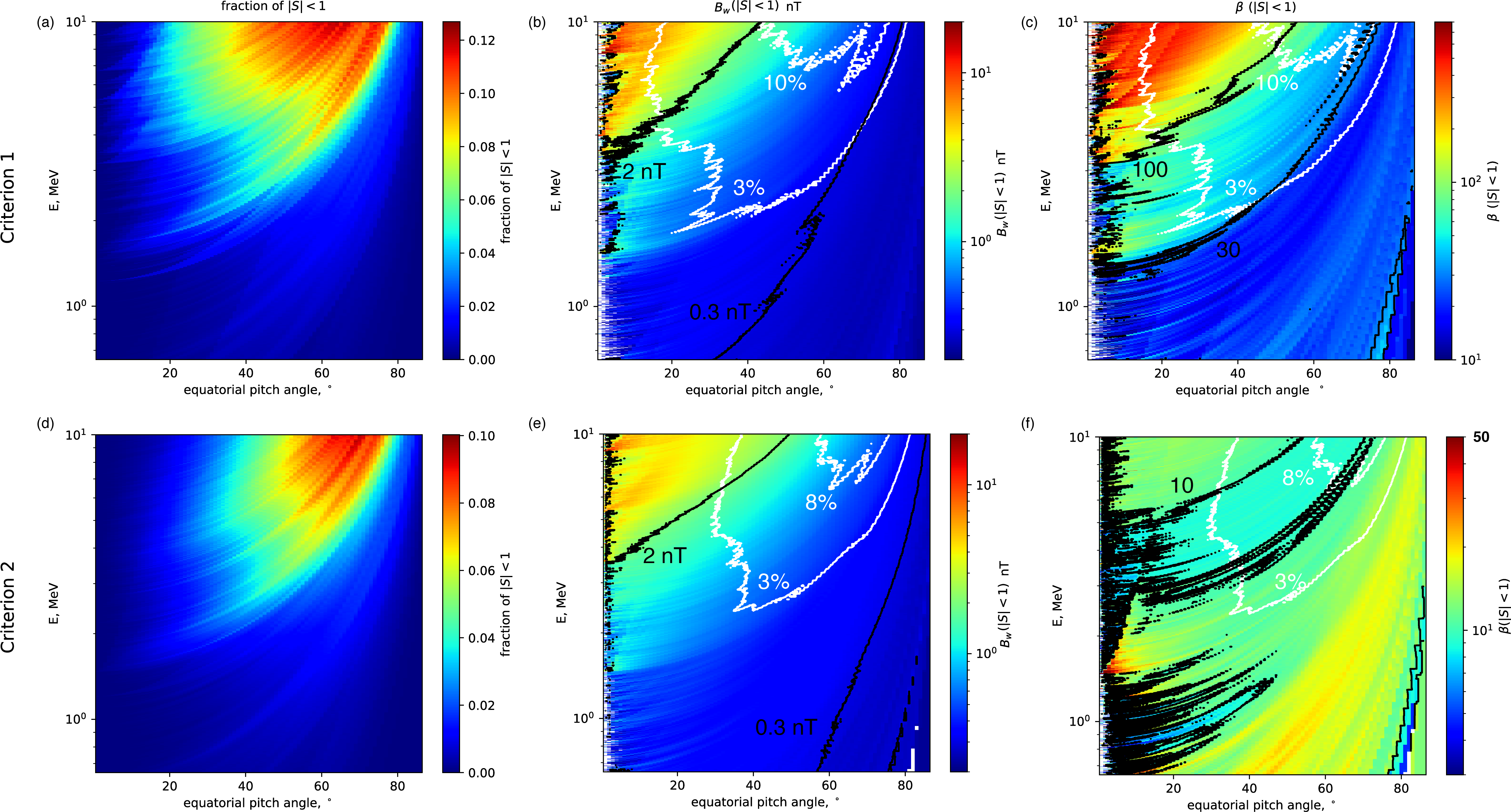}
\end{tabular}
\caption{(a) Fraction of H-band EMIC wave-packets with $|S|<1$ as a function of electron energy $E$ and equatorial pitch-angle $\alpha_{eq}$, using Criterion 1. (b) Average peak wave amplitude $B_{w,peak}$ of $|S|<1$ wave-packets based on Criterion 1. (c) Average size $\langle\beta\rangle$ of $|S|<1$ wave-packets using Criterion 1. Contours in (b,c) show levels of the distribution from (a). (d,e,f) Same as (a,b,c) based on Criterion 2. }
\label{fig3}
\end{figure*}

Figure \ref{fig3}(a,d) show for Criterion 1 and 2, respectively, the average value of the fraction of H-band EMIC wave packets reaching the threshold $|S|<1$ for nonlinear wave-particle interaction, as a function of electron equatorial pitch-angle $\alpha_{eq}$ and energy $E$. Note that this fraction is weighted by packet duration, i.e., it is the ratio of the duration of wave packets with $|S|<1$ to the total time of wave-packet measurements. Figure \ref{fig3}(b,c) shows the average amplitude of waves with $|S|<1$ in the same format, with the contours transferred from Figure \ref{fig3}(a,d). Figure \ref{fig3}(c,f) shows the wave-packet length $\beta$, in the same format as Figure \ref{fig3}(b,c). Based on Criterion 1, most packets allowing nonlinear interaction are relatively long to very long, with average sizes $\beta=30-500$. The fraction of wave packets reaching the threshold for nonlinear interaction varies between $3$\% and $10$\% and corresponds to peak amplitudes $B_{w,peak}\sim0.3-2$ nT. Noticeably, a significant fraction ($>10\%$) of EMIC wave packets, mostly with amplitudes larger than $1$ nT, are able to interact nonlinearly with $>2$ MeV electrons around $\alpha_{eq}\sim 40^{\circ}-60^{\circ}$. However, using Criterion 2 shows that most of these packets with $|S|<1$ contain a strong inner modulation of $B_w$ by at least a factor of $2$ around each peak, $B_{w,peak}$, corresponding to a shorter average packet size $\beta\sim 10-15$. Such a strong wave amplitude modulation may reduce the effectiveness (duration) of nonlinear electron trapping, especially in the presence of simultaneous frequency and phase jumps in-between packets \cite{Usanova10, Nakamura15:emic, Liu18:emic, Grach21:emic}, as in the case of realistic short chorus wave packets \cite{Tao13, Zhang20:grl:phase, An22:Tao}.

\section{Wave frequency variations}\label{sec:frequency}
\begin{figure*}
\centering
\includegraphics[width=1\textwidth]{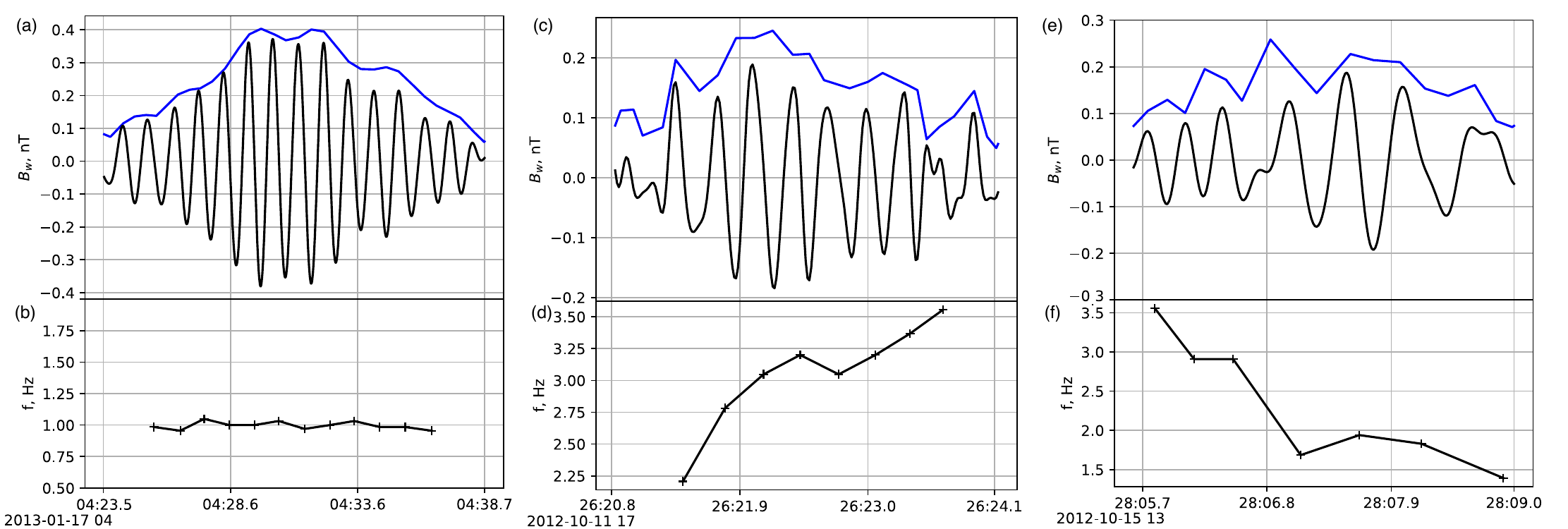}
\caption{Three examples of wave packets with $f\approx const$ (a), $\partial f/\partial t>0$ (b), and $\partial f/\partial t<0$ (c).}
\label{fig4}
\end{figure*}

\begin{figure*}
\centering
\includegraphics[width=0.8\textwidth]{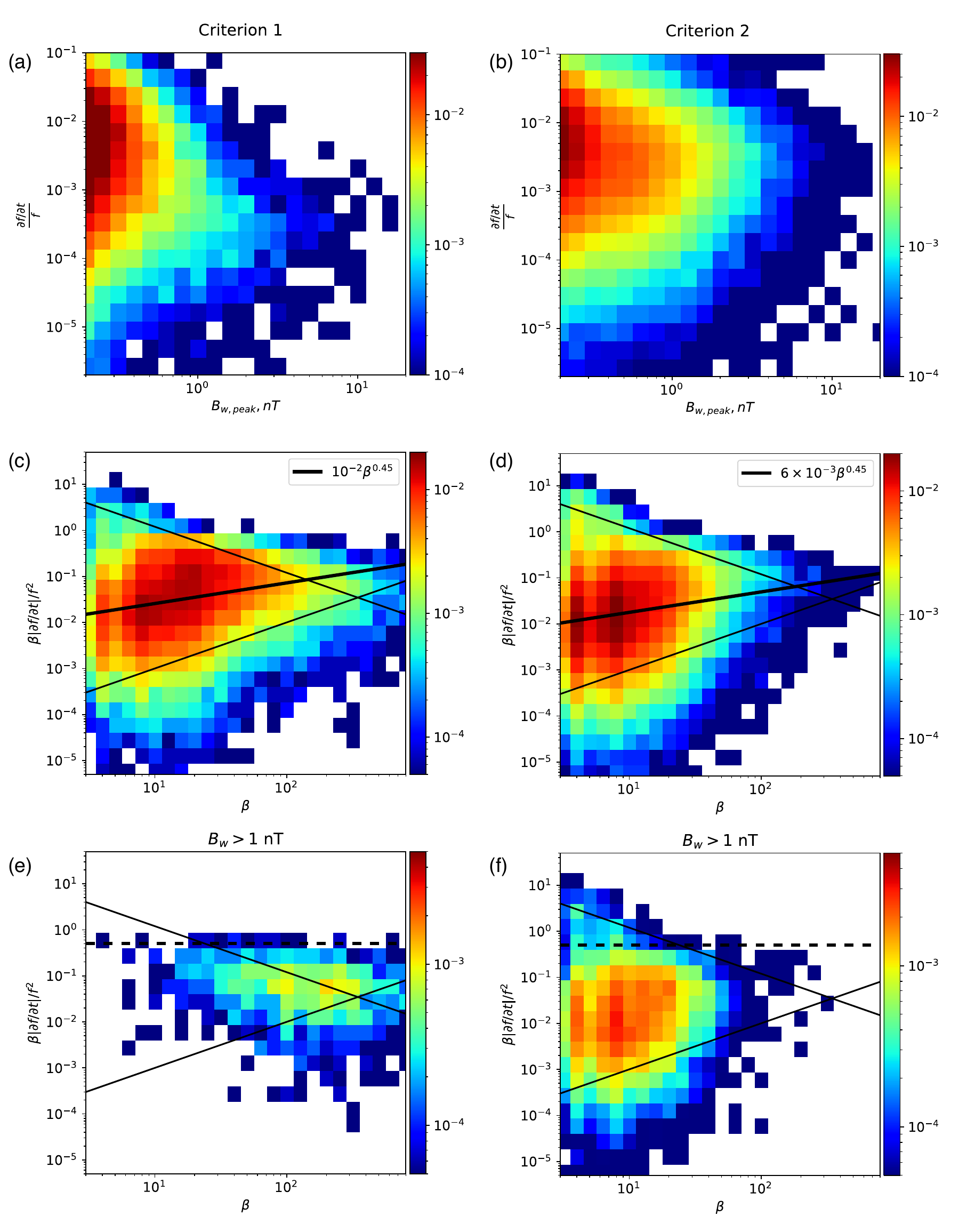}
\caption{Distribution of observed H-band EMIC wave packets in $(B_{w,peak}, \partial f/\partial t/f)$ space for criteria 1 (a) and 2 (b), where $\partial f/\partial t$ is determined by linear fitting of the middle portion of each wave-packet. (c,d) Same as (a,b) for the distribution of packets in $\left[\beta, \beta(\partial f/\partial t/f)/f^2\right]$ space. The lowest and highest thin solid black lines show two theoretical scalings based on the nonlinear theory of EMIC wave growth and wave superposition, respectively (see text). The thick solid black line shows the least-squares power-law fit to the full distribution. (e,f) Same as (c,d) for intense packets with $B_{w,peak}>1$ nT. A dashed black line shows the scaling corresponding to the maximum frequency sweep rates due to the limited frequency range of EMIC waves. }
\label{fig5}
\end{figure*}

An important characteristic of wave-packets is the wave frequency variation within a packet, because this variation may destroy nonlinear resonant wave-particle interaction, thereby significantly reducing the efficiency of nonlinear phase trapping \cite<see discussion in>{Zhang20:grl:frequency,Zhang20:grl:phase,An22:Tao}. Figure \ref{fig4} shows three examples of wave-packets with nearly constant frequency, rising frequency ($\partial f/\partial t>0$) and falling frequency ($\partial f/\partial t<0$). The frequency change is determined by linear regression of the half-wave periods obtained from two successive zero points of the transverse component of the wave field. We calculated $\partial f/\partial t$ inside each wave-packet. The wave-packet distribution in $(\beta,\partial f/\partial t)$ space for the two Criteria of wave-packet selection are shown in Figure \ref{fig5}. Because there are often frequency jumps near the edges of packets, we calculated the linear regression only for the center part of each wave-packet, i.e., where $B_w$ is above the $25^{th}$ percentile of the amplitudes inside the packet. 

As seen in Figures \ref{fig5}(a,b), obtained using Criteria 1 and 2, respectively, most of the H-band EMIC wave packets have a frequency sweep rate $\partial f/\partial t\approx 10^{-3}-10^{-1}$ Hz/s, with no clear dependence on the peak packet amplitude $B_{w,peak}$. This is similar to the range of EMIC frequency sweep rates obtained in previous case studies \cite{Nakamura15:emic, Nakamura19:emic}. The range of measured sweep rates in our statistics agrees with typical sweep rates derived from the nonlinear theory of EMIC wave growth for realistic wave amplitude and plasma parameters \cite{Omura10}. Note, however, that the frequency within a given observed packet can often show significant oscillations, even within relatively short packets (e.g., see Figure \ref{fig4}(d,f)).

Figures \ref{fig5}(c,d) show that most EMIC wave packets are clustered between two thin solid black lines, regardless of the criterion (1 or 2) used to determine packet size. The lowest thin solid black line in Figures \ref{fig5}(c-f) shows a normalized frequency sweep rate $(\partial f/\partial t)/f^2 = 10^{-4}$, independent of $\beta$. It shows low or moderate frequency sweep rates, which increase linearly with $B_{w,peak}$ in Figure \ref{fig5}(a,b), as expected from the nonlinear theory of EMIC wave growth \cite{Omura10, Shoji18}. The highest thin black line in Figures \ref{fig5}(c-f) shows a normalized frequency sweep rate $(\partial f/\partial t)/f^2 = 12/\beta^2$ decreasing as the packet size $\beta$ increases. It represents the typical scaling of the average frequency sweep rate with $\beta$, produced by superposition of two waves of slowly varying amplitude and frequency, as in the case of whistler-mode chorus wave packets \cite<see model details in>{Zhang20:grl:frequency}. Based on the observed trends in the distribution of sweep rates displayed in Figures \ref{fig5}(c,d), short packets with $\beta <50$ and high frequency sweep rate, such that $3/\beta^2 < (\partial f/\partial t)/f^2 < 30/\beta^2$, are probably mainly formed by such wave superposition around their peak power (where $B_{w}(t)>B_{w,peak}/2$). In contrast, longer packets with $\beta>50$ and/or packets with a lower sweep rate $(\partial f/\partial t)/f^2 < 1/\beta^2$ are probably mainly formed by nonlinear trapping-induced amplitude modulation during wave growth \cite{Shoji&Omura13, Nakamura15:emic, Tao17:generation, Mourenas22:jgr:ELFIN}, without significant superposition. The distribution of packet occurrences in Figures 5(c,d) exhibits an intermediate trend due to the interplay of different sweep rates. Low to moderate nonlinear sweep rates remain relatively independent of packet size. In contrast, high sweep rates, which arise from wave superposition, decrease as the packet size increases. This combination of effects results in the intermediate trend observed between the two thin black curves in the figures. The least-squares power-law fit of this global trend is $(\partial f/\partial t)/f^2 \approx 10^{-2}\times \beta^{-0.55}$ (shown by a thick black line). This global trend is close to the expected trend due to nonlinear trapping-induced modulation (with $(\partial f/\partial t)/f^2 \propto const$) than to the trend corresponding to wave superposition (with $(\partial f/\partial t)/f^2 \propto \beta^{-2}$). This suggests a dominant role of trapping-induced amplitude modulation in the formation of EMIC wave packets, contrary to the case of short chorus wave packets \cite{Zhang20:grl:frequency}.

The decreasing trend towards larger $\beta$ in the distribution of the highest sweep rates of the most intense packets (with $B_{w,peak}>1$ nT) in Figure \ref{fig5}(f), determined using Criterion 2, probably means that such packets (or sub-packets) are mainly formed by wave superposition near their peak power. However, the increasing trend of the distribution of packets with low to moderate $\beta(\partial f/\partial t)/f^2$ as $\beta$ increases in Figure \ref{fig5}(f), suggests that the majority of these packets are formed by nonlinear modulation. Using Criterion 1 in Figure \ref{fig5}(e) to examine the same packets with $B_{w,peak}>1$ nT over their whole length down to $0.2$ nT, shows that they are very long ($\beta = 50-1000$), and that their normalized sweep rate over this full length becomes nearly independent of $\beta$, with a maximum scaling $\beta(\partial f/\partial t)/f^2 \approx 0.5$ independent of $\beta$ (see dashed black line). The maximum sweep rate of long and intense lower-band chorus wave packets has a similar scaling, $(\partial f/\partial t)/f^2 \sim 1/\beta$, which stems from the limited frequency bandwidth of lower-band chorus waves \cite{Teng17, Zhang20:grl:frequency}. Similarly, the maximum normalized frequency bandwidth of intense H-band EMIC waves is typically limited to $\Delta f/f_{cp}\sim 0.2$ (with an average frequency $\langle f\rangle/f_{cp}\sim0.4$ at peak power), probably due to both strong damping as $f$ increases toward the proton gyrofrequency $f_{cp}$ and a narrow stop band above the helium gyrofrequency at $f_{cp}/4$ in a realistic plasma composed of protons with a small fraction of helium ions \cite{Kersten14, Ross21}. Since the duration of these packets is $\Delta t = \beta/\langle f\rangle$, this gives an upper limit $\max(\beta\,(\partial f/\partial t)/f^2) = \beta\,(\Delta f/\Delta t)/\langle f\rangle^2 \sim 0.2\,f_{cp}/\langle f\rangle \sim 0.5$ for the longest packets with $\beta >50$ and high sweep rates. The scaling of this theoretical upper limit (shown by a dashed black line) agrees well with the maximum observed values of $\beta\,(\partial f/\partial t)/f^2$ for long ($\beta>50$) and intense ($B_{w,peak}>1$ nT) EMIC wave packets in Figures \ref{fig5}(e,f).

\begin{figure*}
\centering
\includegraphics[width=0.95\textwidth]{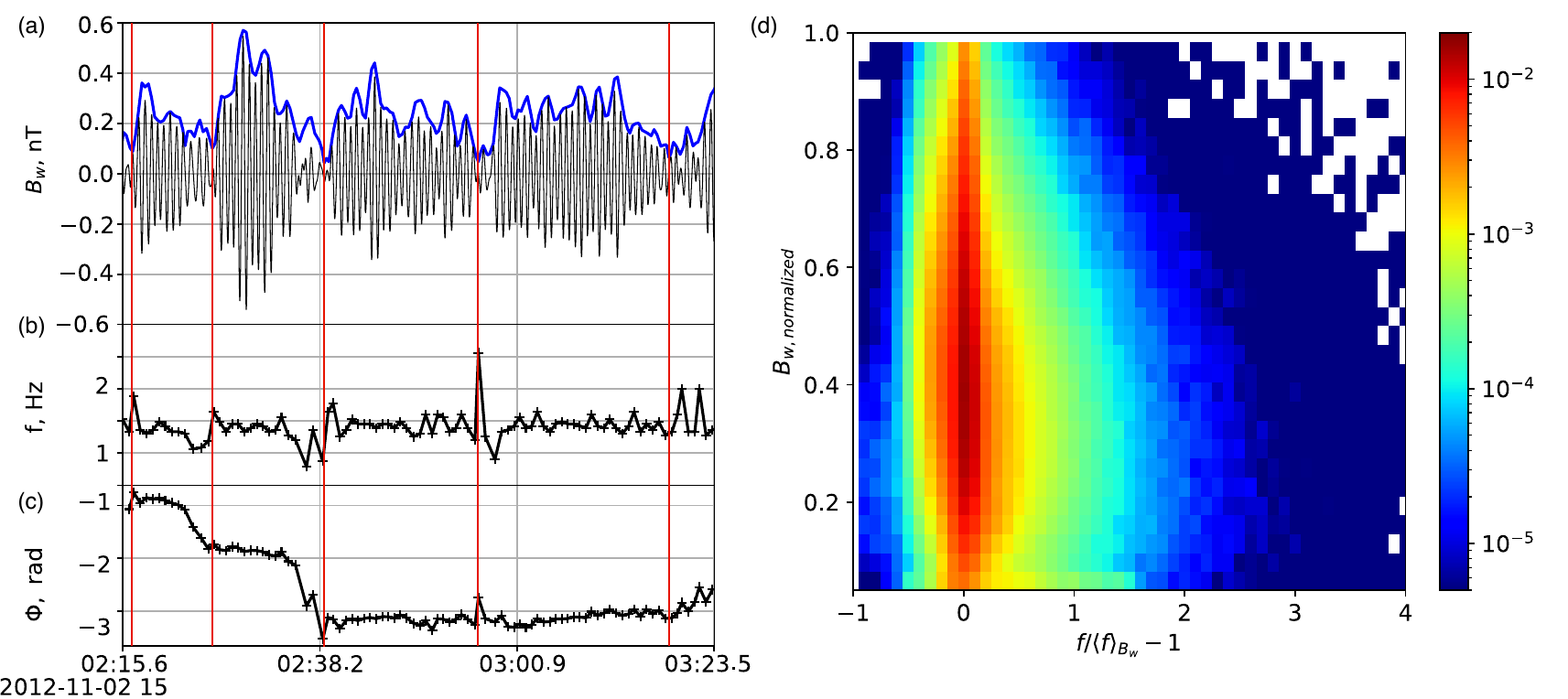}
\caption{(a,b,c) Amplitude, frequency, and wave-normal angle variations of examples of H-band EMIC wave packets. (d) Distribution of H-band EMIC wave packets in $(f/\langle f \rangle_{B_w}-1, B_w/B_{w, max})$ space for all wave packets in out database. $B_{w, max}$ is the maximum amplitude for each wave-packet.}
\label{fig6}
\end{figure*}

Significant, random frequency jumps are observed near the edge of EMIC wave packets. Figures \ref{fig6}(a,b,c) provide examples of fast frequency increases or decreases (with only weak simultaneous wave-normal angle $\Phi$ variations) at the boundary of several packets, where the wave amplitude is small. We use $f/\langle f \rangle_{B_w}-1$ to describe the distribution of frequency variations inside each wave packet where $\langle f \rangle_{B_w}$ is the average frequency weighted by wave amplitude. Figure \ref{fig6}(d) shows that the distribution of relative frequency variations $f/\langle f \rangle_{B_w}-1$ is quite wide, with $\sim10$\% of the wave frequencies reaching $f\sim0.6\langle f \rangle_{B_w}$ and $\sim0.2$\% reaching $f\sim2\langle f \rangle_{B_w}$. Similar results have been obtained for the distribution of frequency variations inside chorus wave packets \cite{Zhang20:grl:frequency}, suggesting the presence of similar physical phenomena in both EMIC and chorus wave packets, such as wave superposition and nonlinear amplitude modulation.

\section{Implications for Electron Precipitation by Intense H-band EMIC Wave Packets}\label{sec:Implications}

In this section, we use our statistics of H-band EMIC wave packets to determine the main (MLT,$\Omega_{pe}/\Omega_{ce}$) ranges of high-energy electron precipitation by such intense packets, and the relative importance of quasi-linear, nonlinear, and so-called nonresonant interactions for electron precipitation at high and low energy.

\begin{figure*}
\centering
\includegraphics[width=1\textwidth]{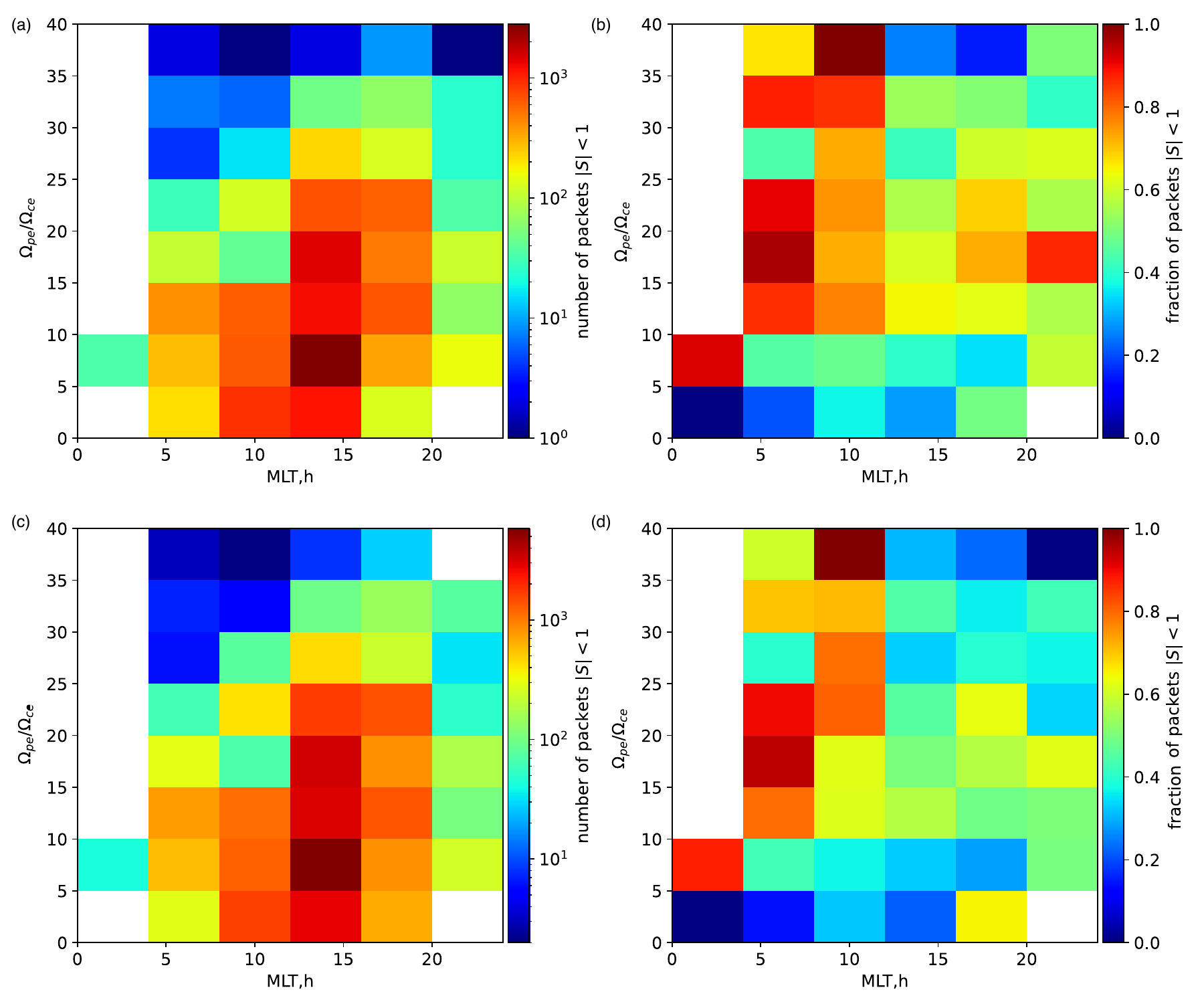}
\caption{Number of H-band EMIC wave packets (a,c) and fraction of H-band EMIC wave packets (b,d) with $|S|<1$ for $E=3-10$ MeV and $\alpha_{eq}=40^\circ-60^\circ$ in (MLT, $\Omega_{pe}/\Omega_{ce}$) space, for criteria 1 (a,b) and 2 (c,d). Note that the gap in MLT=0-4 is due to the absence of events.}
\label{fig7}
\end{figure*}

Figure \ref{fig7} shows that the fraction of H-band EMIC wave packets with $|S|<1$ for $E=3-10$ MeV and $\alpha_{eq}=40^\circ-60^\circ$ is larger when $\Omega_{pe}/\Omega_{ce}>10$ than when $\Omega_{pe}/\Omega_{ce}<10$ ($\sim50-80$\% versus $\sim20-50$\%), but also sensibly larger at $4-11$ MLT ($\sim70-80$\%) than at 12-24 MLT ($\sim50-60$\%). This implies that nonlinear interactions should be on average more important at $4-11$ MLT than at $12-24$ MLT for electrons within these energy and pitch-angle ranges, which are the most likely to experience nonlinear trapping for typical wave and plasma parameters \cite{Omura&Zhao12, Kubota&Omura17, Grach22:elfin}. However, Figures \ref{fig7}(b,d) suggest that nonlinear interactions probably play a significant role in the precipitation of $>3$ MeV electrons by H-band EMIC waves when $\Omega_{pe}/\Omega_{ce}>10$ over the whole $5-24$ MLT domain, through an initial nonlinear trapping at $\alpha_{eq}=40^\circ-60^\circ$ by the highest amplitude part of an intense wave packet, leading to a rapid decrease of $\alpha_{eq}$, followed by scattering into the loss-cone by the lower amplitude part of this packet or other low amplitude packets \cite{Omura&Zhao12, Kubota&Omura17, Nakamura19:emic, Grach22:elfin}. Figures \ref{fig7}(a,c) also show that the absolute number of wave packets with $|S|<1$ is larger at $9-20$ MLT and when $\Omega_{pe}/\Omega_{ce}<25$. This is consistent with the lower occurrence rate of all H-band EMIC waves and their reduced average intensity at $0-8$ MLT compared with $8-20$ MLT and for $\Omega_{pe}/\Omega_{ce}>25$ compared with $\Omega_{pe}/\Omega_{ce}<25$ \cite{Ross21}.

\begin{figure*}
\centering
\includegraphics[width=1.\textwidth]{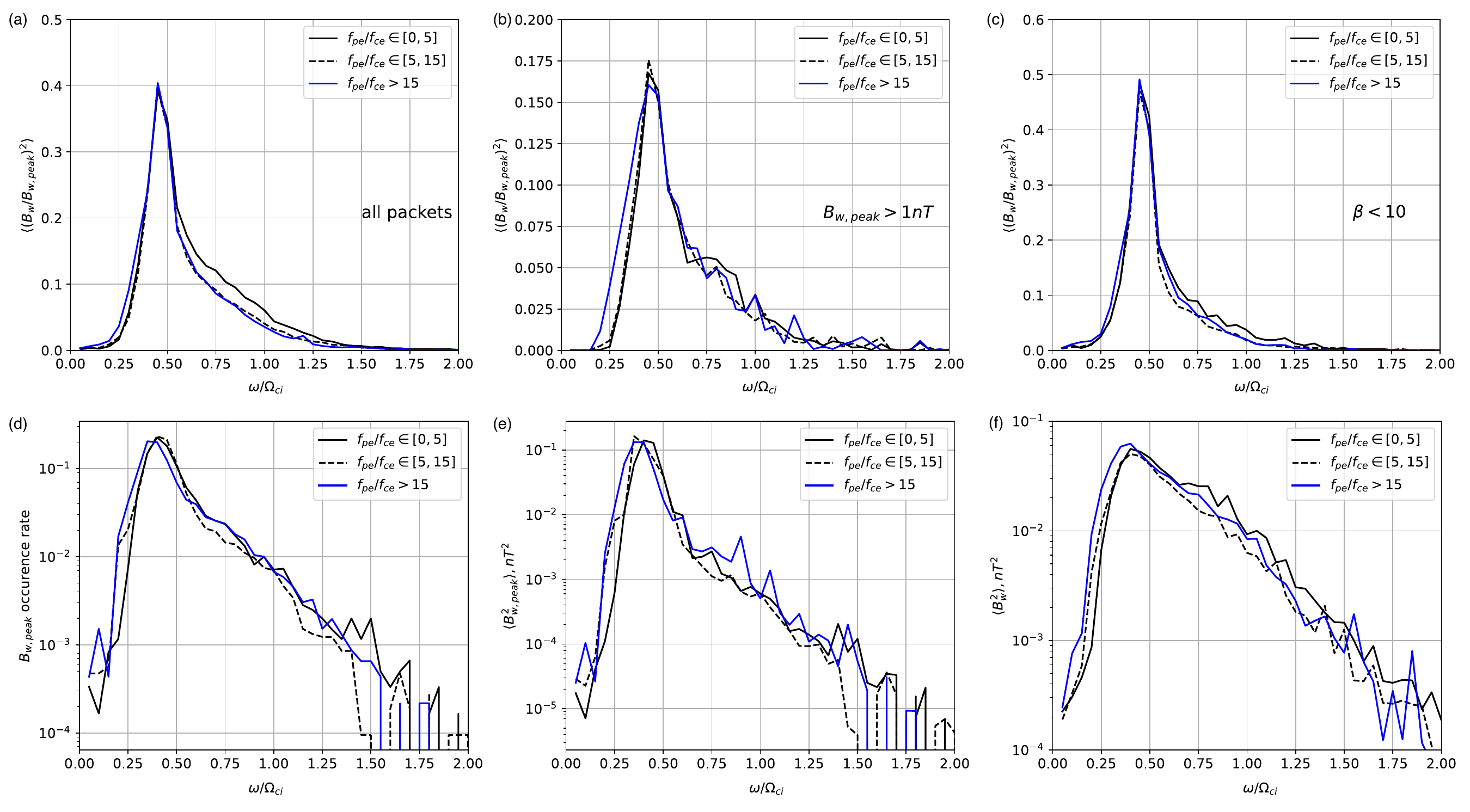}
\caption{Statistical distribution of $\langle B_w^2(\omega/\Omega_{ci})/B_{w,peak}^2\rangle$  as a function of $\omega/\Omega_{ci}$ inside H-band EMIC wave packets (determined using Criterion 1) in the 10-16 MLT sector, such that their peak amplitude $B_{w,peak}$ is reached at $\omega/\Omega_{ci}\in[0.40,0.50]$, for three $\Omega_{pe}/\Omega_{ce}$ ranges ($0-5$, $5-15$, $>15$). (b,c) Same as (a) for all packets determined using Criterion 1 in the 10-16 MLT sector with $B_{w,peak}>1$ nT; and for the fraction of packets with $\beta<10$, respectively. (d,e,f) The occurrence rate of the peak amplitude of a packet as a function of $\omega/\Omega_{ci}$; $\langle B_{w,peak}^2(\omega/\Omega_{ci})\rangle$; and $\langle B_{w}^2(\omega/\Omega_{ci})\rangle$, respectively, for all packets determined using Criterion 1 in the 10-16 MLT sector, for the same $\Omega_{pe}/\Omega_{ce}$ ranges. }
\label{fig8}
\end{figure*}

Figure \ref{fig8}(a) shows the statistical distribution of the average of the normalized wave power $\langle B_w^2(\omega/\Omega_{ci})/B_{w,peak}^2\rangle$ inside H-band EMIC wave packets (determined using Criterion 1) in the 10-16 MLT sector where low energy electron precipitation is most effective \cite{Angelopoulos23:SSR}, for three $\Omega_{pe}/\Omega_{ce}$ ranges ($0-5$, $5-15$, $>15$). We only kept wave packets determined based on Criterion 1, with a peak amplitude $B_{w,peak}>200$ pT occurring at $\omega/\Omega_{ci}\in[0.4,0.5]$ where the most intense H-band waves are statistically observed \cite{Zhang16:grl}. For each packet, we calculated $B_w(\omega/\Omega_{ci})$ within the packet (during each half wave period), as well as the maximum frequency $\omega_{max}$ reached within this packet, and a null amplitude, $B_w(\omega/\Omega_{ci})=0$, was assigned to frequencies $\omega>\omega_{max}$ not present within the packet. Finally, an averaging was performed over all these packets. 

Figures \ref{fig8}(a,b,c) show: the average wave power spectrum of typical intense packets; of the most intense packets with $B_{w,peak}>1$ nT; and of the shortest packets with $\beta<10$. The wave power within these packets remains surprisingly elevated up to high frequencies, reaching on average $\sim10-25$\% of their peak power $B_{w,peak}^2$ at $\omega/\Omega_{ci}=0.7-0.9$, with only slightly higher or lower levels in the most intense or shortest packets. Taking into account the smaller dataset available at 0-6 MLT, there is no significant dependence on the electron plasma frequency to gyrofrequency ratio $\Omega_{pe}/\Omega_{ce}$. Since sub-relativistic ($<0.5$ MeV) electron precipitation through cyclotron resonance with H-band EMIC waves requires a high ratio $\Omega_{pe}/\Omega_{ce}>10-15$ and a high frequency $\omega/\Omega_{ci}>0.7$ \cite{Summers&Thorne03}, these results imply that cyclotron resonance with intense H-band EMIC wave packets in high-density regions likely plays an important role in the sub-relativistic electron precipitation frequently observed during conjunctions with EMIC wave bursts measured near the equator \cite{Hendry17, Hendry19, Capannolo19, Zhang21, Angelopoulos23:SSR}, whatever their size $\beta$ and peak amplitude. 

However, Figures \ref{fig8}(b,c) indicate that the most intense and longest packets are comparatively more likely to produce sub-relativistic electron precipitation through cyclotron resonance. Although this may suggest that nonlinear interactions (trapping) with intense packets could be important for sub-relativistic electron precipitation, it is worth noting that only a small fraction, $\sim3-8$\%, of the packets can reach the threshold $|S|<1$ for nonlinear interaction with $0.3-0.8$ MeV electrons in Figure \ref{fig3}, only at $\alpha_0>40^\circ$. Moreover, Figure \ref{fig8}(f) shows that inside all packets at 10-16 MLT, the average power $\langle B_w^2\rangle$ of waves present at $\omega/\Omega_{ci}\sim0.7$ when $\Omega_{pe}/\Omega_{ce}>15$ corresponds to low amplitudes $B_w\sim140$ pT that may be insufficient for trapping. Therefore, the most probable origin of sub-relativistic electron precipitation is electron scattering near the loss-cone \cite{Kubota&Omura17} through cyclotron resonance with the higher-frequency, lower-power part of intense H-band EMIC wave packets. A similar conclusion has recently been drawn from the good agreement, over a wide energy range ($0.2-1.5$ MeV), between average precipitating to trapped electron flux ratios measured by the low-altitude ELFIN CubeSats at $L\sim5-6$ in the noon-dusk sector and flux ratios inferred from electron quasi-linear diffusion through cyclotron resonance with statistical H-band EMIC wave power spectra measured by the Van Allen Probes \cite{Angelopoulos23:SSR}, although that study could not identify the dominant role of intense packets.

Interestingly, the shortest packets, with $\beta<10$, also comprise a significant wave power at high frequencies, which may be surprising at first sight since all packets selected in Figures \ref{fig8}(a,b,c) reach their maximum power at $\omega/\Omega_{ci}=0.4-0.5$. But Figure \ref{fig5} shows that short packets can have high frequency sweep rates $\partial f/\partial t>0.05$ Hz/s, sufficient to reach such high frequencies over less than 10 wave periods. Figure \ref{fig6}(d) also shows that frequency oscillations within packets often reach $\sim1.6-2$ times their average frequency, consistent with results in Figure \ref{fig8}(c). Figure \ref{fig8}(c) indicates that short EMIC wave packets, which may be more capable of producing nonresonant scattering of sub-relativistic electrons \cite{Chen16:nonresonant, An22:prl}, similarly contain a significant fraction of average wave power at high frequencies $\omega/\Omega_{ci}\sim0.7$, much higher than the frequency $\omega/\Omega_{ci}\sim0.4-0.5$ of the main waves at peak power. This indeed allows sub-relativistic electron scattering via cyclotron resonance with the high-frequency, high-$k$ tail of the EMIC wave power spectrum, as discussed earlier. Accordingly, EMIC wave-packet statistics presented in Figure \ref{fig8} suggest that resonant scattering by the upper-frequency tail of the wave power spectrum of intense packets may be sufficiently frequent to account for most events of sub-relativistic electron precipitation.

It is worth emphasizing again here that high $\omega/\Omega_{ci}>0.7$ H-band EMIC waves naturally correspond, through their dispersion relation, to high wave numbers $k$ \cite{Summers&Thorne03, Denton19}, allowing cyclotron resonance with electrons of lower energy than in the case of more typical lower-frequency waves (e.g., see section \ref{Sec:NR}). Basically, a high frequency corresponds to a short time interval $\sim\pi/\omega$ between two successive changes of sign of the wave electric field, which should translate in space into a short distance $\sim\pi/k$ between similar changes of sign, with $k = \omega \times 1/v_{ph}$ and $v_{ph}$ the wave phase velocity for EMIC waves obeying the dispersion relation. Even for high $\omega/\Omega_{ci}$ waves produced by wave superposition, one still expects to get $k \sim \omega \times (1/v_{ph}$ to $1/v_g)$, where the parallel group velocity $v_g$ is lower than $v_{ph}$ for $\omega/\Omega_{ci}>0.5$, giving $k$ values at least as high as based on the dispersion relation. From another perspective, the $k$ distribution due to the superposition of two or more waves of slowly varying amplitudes is expected to be similar to the distribution of $\omega/\Omega_{ci}$ values due to superposition, with a similar heavy tail up to at least $\sim 2\langle k\rangle$ \cite{Zhang20:grl:frequency}. For $\langle\omega/\Omega_{ci}\rangle\sim0.45$, $k\sim 2\langle k\rangle$ corresponds to $\omega/\Omega_{ci}\sim0.7$ via the dispersion relation. This suggests that most waves with $\omega/\Omega_{ci}\sim0.7-0.8$ in Figures \ref{fig8}(a,b,c) should indeed reach cyclotron resonance with sub-relativistic electrons.

Figures \ref{fig8}(d,e) show the occurrence rate of the peak amplitude of a packet and the average peak power $\langle B_{w,peak}^2\rangle$ (weighted by the occurrence rate of peak power in each normalized frequency bin) as a function of $\omega/\Omega_{ci}$, for all packets at 10-16 MLT selected using Criterion 1. While most packets ($\sim 60$\%) reach their peak power $B_{w,peak}^2>0.1$ nT$^2$ at moderate frequencies $\omega/\Omega_{ci}\sim0.35-0.5$, there is also a finite fraction ($\sim 6$\%) of intense packets reaching a peak amplitude $B_{w,peak}\sim 200$ pT at $\omega/\Omega_{ci}\sim0.75-0.90$ when $\Omega_{pe}/\Omega_{ce}>15$, corresponding to a much lower $\langle B_{w,peak}^2\rangle\sim0.0025$ nT$^2$. Finally, Figure \ref{fig8}(f) shows the average power $\langle B_w^2\rangle$ of H-band EMIC waves from intense packets at 10-16 MLT that are present in each $\omega/\Omega_{ci}$ bin. Waves at high frequencies $\omega/\Omega_{ci}\sim0.7-0.8$ still reach root-mean-square amplitudes of $\sim100-140$ pT. 

The wave power distributions provided in Figures \ref{fig8}(a,b) could be used to calculate the ratio of quasi-linear diffusion rates at low and high energy from the subset of EMIC waves within such intense packets, for comparisons with events of rapid electron loss, where precipitation at low energy $0.2-0.7$ MeV is much less efficient than above $2$ MeV but still discernible in low-altitude spacecraft observations during conjunctions with EMIC wave bursts measured near the equator \cite{Hendry17, Capannolo19, Zhang21, Angelopoulos23:SSR,Capannolo23}.

\section{Discussion and Conclusions}\label{sec:conclusions}

\begin{figure*}
\centering
\includegraphics[width=1\textwidth]{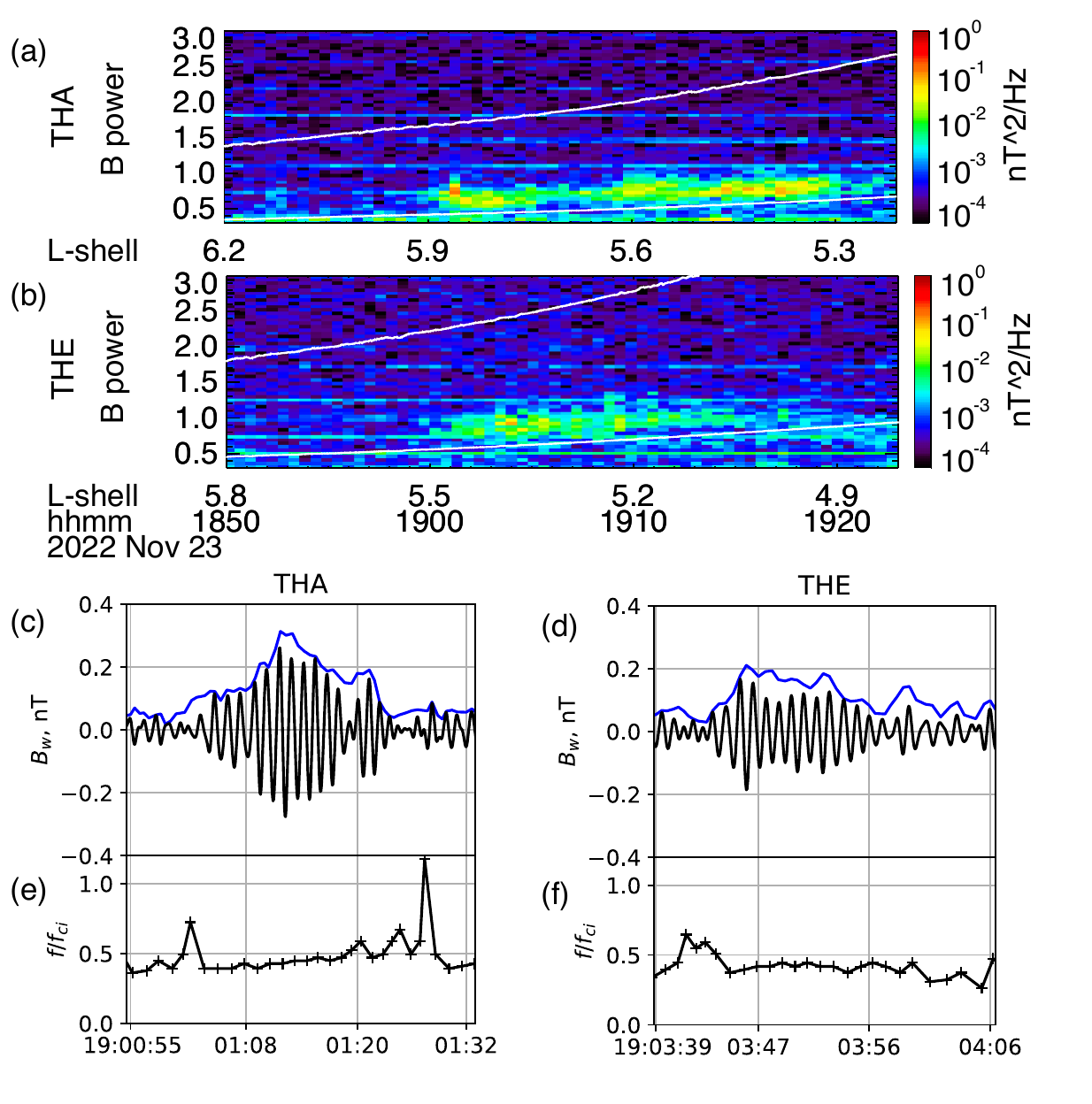}
\caption{The wave power spectrum of $H$ band EMIC waves observed by THEMIS-A and THEMIS-E (a, b). Two white lines from top to bottom indicate the proton and helium cyclotron frequency, respectively. Two examples of wave packets, showing their amplitude and frequency variations, from THEMIS-A and THEMIS-E, respectively (c-f). We use magnetic field measurements with $1/5$ s sampling rate ($fgl$ dataset) from the Fluxgate magnetometer.}
\label{fig9}
\end{figure*}

Although the present results are based on statistics of individual EMIC wave-packets, EMIC wave characteristics are expected to remain roughly similar within the entire EMIC wave source region, of typical spatial extent $\sim 0.5\,R_E$ to $\sim 1\,R_E$ at the equator \cite{Blum16,Blum17}. Therefore, EMIC wave-packet characteristics, specific inside each wave source region, are expected to determine the properties of the associated electron precipitation over a relatively large spatial domain, much larger than the domains of precipitation bursts driven by individual whistler-mode wave packets \cite<see discussion in>{Zhang23:jgr:ELFIN&scales}. Figure \ref{fig9} shows an example of simultaneous EMIC wave-packet observations by two THEMIS spacecraft \cite{Angelopoulos08:ssr} with $\sim 1\,R_E$ azimuthal separation: although wave spectra and individual wave-packets show some differences, the general wave-packet characteristics (number of wave periods, average frequency, and frequency jumps at the wave-packet edges) remain similar. Further investigations of multi-spacecraft missions will be needed to statistically confirm this expected homogeneity of EMIC wave-packet characteristics within a given wave source region. In addition, the spatial scale of the region occupied by EMIC waves increases during their propagation to middle latitudes \cite{Kim&Johnson16,Hanzelka23:emic_propagation}. This should further increase the spatial extent of electron precipitation driven by similar EMIC wave-packets. 

In this study, we used 3 years of Van Allen Probe observations to provide the distributions of wave amplitudes, wave-packet sizes, and rates of frequency variations within individual intense H-band EMIC wave-packets. We found that most of such intense wave-packets are short, with $\sim10$ wave periods each, and that up to 10\% of such packets can attain amplitudes that enable nonlinear resonant interaction with multi-MeV electrons. Up to 3\% of these packets can reach nonlinear resonant interaction with 2 MeV electrons, mainly for equatorial pitch-angles $>20^\circ-30^\circ$. Frequency variations within packets often reach $\sim50-100$\%. We showed that such wave packet characteristics are likely mainly due to the presence of amplitude modulations due to nonlinear trapping during EMIC wave growth \cite{Shoji&Omura13, Nakamura15:emic}, although a significant fraction of the packets (especially packets with high frequency sweep rates) probably result from wave superposition. In comparison, short chorus wave packets mostly result from wave superposition \cite{Zhang20:grl:frequency, Zhang21:jgr:data&model, Nunn21}. The range of observed frequency sweep rates within packets agrees well with the nonlinear theory of EMIC wave growth \cite{Omura10}.

We examined the implication of H-band EMIC wave packet characteristics for electron precipitation and the regime of wave-particle interactions. The occurrence rate of intense wave packets potentially reaching the nonlinear regime (with $|S|<1$ for $E=3-10$ MeV and $\alpha_{eq}=40^\circ-60^\circ$) was found to have a broad maximum at 5-24 MLT and at $\Omega_{pe}/\Omega_{ce}>10$, suggesting an important role of such intense packets in rapidly transporting multi-MeV electrons to lower pitch-angles $\alpha_{eq}=10^\circ-20^\circ$ through trapping, before a subsequent scattering eventually leads to their precipitation into the atmosphere \cite{Kubota&Omura17}. 

We have also shown that the average wave power spectrum of intense H-band EMIC wave packets contains a significant high-frequency tail, reaching $\sim10-25$\% of their peak power at $\omega/\Omega_{ci}=0.7-0.9$, with a more (less) substantial high-frequency tail inside the most intense (the shortest) packets and nearly no dependence on the $\Omega_{pe}/\Omega_{ce}$ range ($0-5$, $5-15$, or $>15$). Sub-relativistic ($<0.5$ MeV) electron precipitation through cyclotron resonance with such waves can occur for $\omega/\Omega_{ci}>0.7$ and 
$\Omega_{pe}/\Omega_{ce}>10-15$. \citeA{Angelopoulos23:SSR} have already demonstrated that sub-MeV electron precipitation observed by ELFIN CubeSats in the noon-dusk sector concurrently with stronger EMIC wave-driven $>1$ MeV precipitation has a spectral shape consistent (down to $\sim200-300$ keV) with quasi-linear resonant scattering by the high-frequency tail of the average H-band EMIC wave power spectrum recorded by the Van Allen Probes in this sector \cite{Zhang16:grl}. Here, we have further shown that the fraction of EMIC wave power at such high frequencies is much larger within intense packets than in previous global statistics \cite{Zhang16:grl} that perform an average over all EMIC waves present within and outside intense packets. The typical amplitudes of such high-frequency waves are $\sim100-150$ pT. Therefore, our statistical results suggest that quasi-linear resonant interactions with the high-frequency portion of the power spectrum of intense H-band EMIC wave packets, which exhibit strong amplitude modulations, likely provide the main contribution to the sub-relativistic electron precipitation observed during EMIC wave bursts \cite{Hendry17, Hendry19, Capannolo19,Capannolo23, Zhang21}, in agreement with other recent studies \cite{Denton19, An22:prl, Angelopoulos23:SSR}.

\acknowledgments
We gratefully acknowledge the Van Allen Probes EMFISIS team for providing the data used in this study. A.V.A., X.-J.Z. acknowledge support by NASA awards 80NSSC23K0403, 80NSSC23K0108, and 80NSSC20K1270, and NSF \#2329897. We acknowledge NASA contract NAS5-02099 for the use of data from the THEMIS mission. We thank K. H. Glassmeier, U. Auster, and W. Baumjohann for the use of FGM data provided under the lead of the Technical University of Braunschweig and with financial support through the German Ministry for Economy and Technology and the German Aerospace Center (DLR) under contract 50 OC 0302.

% A.V.A., X.-J.Z., and V.A. acknowledge support by NSF grants AGS-1242918, AGS-2019950, and AGS-2329897. We are grateful to NASA's CubeSat Launch Initiative for ELFIN's successful launch in the desired orbits. We acknowledge early support of ELFIN project by the AFOSR, under its University Nanosat Program, UNP-8 project, contract FA9453-12-D-0285, and by the California Space Grant program. We acknowledge the critical contributions of numerous volunteer ELFIN team student members. We acknowledge support by NSF Magnetospherics Base Program. We thank E.V. Masongsong for editorial assistance with the manuscript.

\section*{Open Research}
\noindent Van Allen Probes wave data is available at \url{https://www.rbsp-ect.lanl.gov/} and \url{https://emfisis.physics.uiowa.edu/data/index}\\
\noindent Data was retrieved and analyzed using SPEDAS V4.1, see \citeA{Angelopoulos19}.
\noindent THEMIS data is available at \url{http://themis.ssl.berkeley.edu.}

% \bibliography{full,addon}

\end{document}